\documentclass{optica-article}
\usepackage{eurosym}
\usepackage{lineno}
\usepackage{graphics}
\usepackage{graphics}
\journal{osajournal}
\articletype{Research Article}
\begin{document}
\title{Effects of fractional diffraction on nonlinear $\mathcal{PT}$ phase
transitions and stability of dark solitons and vortices}
\author{Xueqing He,\authormark{1,2} Mingming Zhang,\authormark{1,2} Pengfei
Li\authormark{1,2,*}, Dumitru Mihalache,\authormark{3} and Boris A. Malomed\authormark{4}}
\address{\authormark{1}Department of Physics, Taiyuan Normal University, Jinzhong, 030619, China\\
\authormark{2}Institute of Computational and Applied Physics, Taiyuan Normal University, Jinzhong, 030619, Shanxi, China\\
\authormark{3}Horia Hulubei National Institute of Physics and Nuclear Engineering, Magurele, Bucharest RO-077125, Romania\\
\authormark{4}Department of Physical Electronics, School of Electrical Engineering, Faculty of Engineering, and Center for Light-Matter Interaction,
Tel Aviv University, Tel Aviv 69978, Israel}
\email{\authormark{*}lpf281888@gmail.com}

\begin{abstract}
The wave propagation under the action of fractional diffraction has recently
drawn increasing attention in nonlinear optics. Here, we address the effect
of fractional diffraction on the existence, phase transitions, and stability
of dark solitons (DSs) and vortices in parity-time ($\mathcal{PT}$)
symmetric graded-index waveguide with self-defocusing nonlinearity. The DSs
and vortices are produced by numerical solution of the corresponding one-
and two-dimensional fractional nonlinear Schr\"{o}dinger equations. We show
that solution branches of fundamental and higher-order DSs collide pair-wise
(merge) and disappear with the increase of the gain-loss strength, revealing
nonlinear $\mathcal{PT}$ phase transitions in the waveguide. Numerically
identifying the merger points, we demonstrate effects of the fractional
diffraction on the phase transition. The phase transition points determine
boundaries of existence regions for the DSs and vortices. The stability of
the DSs and vortices is studied by means of the linearization with respect
to small perturbations. Direct simulations of perturbed evolution
corroborate their stability properties predicted by the analysis of small
perturbations.
\end{abstract}

\section{Introduction}

\label{Sec I} Dark solitons (DSs) feature localized dips on a modulationally
stable continuous wave (or extended finite-width) background \cite{1}. The
DSs are more stable than their bright counterparts against noise
perturbations and are less susceptible to environmental disturbances. Thus,
DSs offer various applications to optical communications and sensing,
and are objects of basic interest to nonlinear optics in general \cite%
{2,3,DSQ}.

The topic of $\mathcal{PT}$ symmetry started in the form of
non-Hermitian quantum mechanics \cite{CMBender,Mostafazadeh,RMPBender},
where it was found that complex potentials obeying the $\mathcal{PT}$
symmetry could exhibit all-real spectra. The increase of the strength of the
imaginary part of the complex potential leads to a phase transition in the
respective linear system at a critical value of the strength, the spectrum
acquiring complex eigenvalues above the critical value. The concept had
later spread out to optics \cite{Christodoulides0,Christodoulides}, where
the $\mathcal{PT}$-symmetry is built as a combination of gain and loss
elements which are mirror images of each other \cite{Segev}. In nonlinear
optics, $\mathcal{PT}$-symmetric solitons are accessible in a wide variety
of nonlinear optical platforms \cite{4,5,6}. In such a setting, dark
solitons were predicted in the framework of the nonlinear Schr\"{o}dinger
equation (NLSE) with the $\mathcal{PT}$-symmetric potential \cite{7}. DSs
also exist in coupled NLSEs, including models of $\mathcal{PT}$-symmetric
dual-core optical couplers \cite{8} (similar to bright solitons predicted in
the same system \cite{Driben,Barash}). In particular, DS dynamics was
studied in $\mathcal{PT}$-symmetric nonlinear directional couplers with
third-order and intermodal dispersions, where phase-controlled switching
dynamics with excellent efficiency was demonstrated with a very low critical
power \cite{9}. Furthermore, the creation of dark solitons and vortex solitons
were proposed in a cold-atom system with linear and nonlinear $\mathcal{PT}$%
-symmetric potentials \cite{STVS0,STVS1,STVS2,STVS3,STVS4,STVS5,STVS6,STVS7,STVS8}.

Recently, optical solitons have been investigated in the framework of
fractional NLSEs (FNLSEs) \cite{10,Chaos,11,12}, which include the effective
fractional diffraction or dispersion represented by the Riesz fractional
derivative with the respective L\'{e}vy index (LI). The linear fractional
Schr\"{o}dinger equation (FSE) originates from fractional-dimensional
quantum mechanics \cite{13,14}. The derivation was based on the Feynman's
path integration, where the superposition of the motions was taken along
paths that correspond not to the usual Brownian random walks, but to
trajectories built by L\'{e}vy flights with the respective LI.

FSE can be emulated in classical optics, as proposed by Longhi \cite{15},
using an optical cavity incorporating two lenses and a phase mask. The
lenses perform direct and inverse Fourier transforms of the light beam
with respect to the transverse coordinate(s), while the phase mask creates
the phase structure corresponding to the action of the fractional
diffraction. Then, the continuous FSE is introduced by averaging over many
cycles of circulation of light in the cavity. Another possibility for the
realization of FSE in optics is provided by the transmission of light pulses
in the temporal domain, with an effective fractional group-velocity
dispersion acting on femtosecond laser pulses \cite{16}.

In the context of optics, it is natural to include nonlinearity
in FSE, leading to fractional NLSE, where the prediction of fractional optical solitons have drawn much interest \cite%
{17,18,19,20,21,22,23,24,25,26,FDVS,FRS}. Furthermore, fractional
diffraction was used to manipulate the Airy beam dynamics \cite{TanC1,TanC2}%
. Earlier, fractional NLSEs with $\mathcal{PT}$-symmetric potentials were
addressed, where symmetric and antisymmetric solitons and ghost states with
complex-conjugate propagation constants were demonstrated \cite{27,28}, as
well as their two-dimensional counterparts \cite{29}.

Previous studies on factional NLSEs focused primarily on properties of
bright solitons, while the existence and stability of non-Hermitian dark
solitons and vortices have not been fully investigated. In this work, we
address fractional DSs and vortices in one- and two-dimensional models of
the nonlinear waveguide with fractional diffraction, characterized by LI
$\alpha $ and the $\mathcal{PT}$-symmetric potential. We identify families
of fractional DSs and vortices, and find branches of these modes that
collide in pairs and disappear at the points of the nonlinear $\mathcal{PT}$
phase transition (which is different from the above-mentioned phase
transition of breaking the real spectrum in linear $\mathcal{PT}$-symmetric
systems). We examine the effect of the fractional diffraction on the phase
transitions and stability of the fractional DSs and vortices, by producing
stationary solutions for them and performing their linear-stability
analysis. We also consider the perturbed evolution of stable and unstable
fractional DSs and vortices, which confirms the predictions of the
linear-stability analysis.

\section{The model and numerical methods}

\label{Sec II}

\subsection{The model}

Our model addresses the propagation of light in the fractional diffraction
medium governed by the dimensionless FNLSE \cite{27,29}

\begin{equation}
i\frac{\partial \Psi }{\partial z}-\frac{1}{2}\left( -\frac{\partial ^{2}}{%
\partial x^{2}}\right) ^{\alpha /2}\Psi +\left[ V\left( x\right) +iW\left(
x\right) \right] \Psi -|\Psi |^{2}\Psi =0,  \label{FNLSE1}
\end{equation}%
where variables $z$ and $x$ represent the normalized propagation length and
coordinate in the transverse direction, respectively, and the normalized
nonlinearity coefficient corresponds to the self-defocusing sign of the
cubic term. The fractional diffraction with LI $\alpha $ is defined as the
above-mentioned Riesz derivative:

\begin{equation}
\left( -\frac{\partial ^{2}}{\partial x^{2}}\right) ^{\alpha /2}u(x)=\frac{1%
}{2\pi }\int_{-\infty }^{+\infty }|p|^{\alpha }dp\int_{-\infty }^{+\infty
}dxe^{ip(x-x^{\prime })}u(x^{\prime }).  \label{Risz derivative}
\end{equation}%
It is built as a nonlocal operator, produced by the juxtaposition of the
direct and inverse Fourier transforms, with fractional differentiation
proper represented by factor $|p|^{\alpha }$ in the Fourier space. Normally,
LI takes values $1\leq \alpha \leq 2$, the limit of $\alpha =2$
corresponding to the canonical (non-fractional) Schr\"{o}dinger equation,
with operator (\ref{Risz derivative}) reducing to the usual second
derivative. Actually, the constraint of $\alpha >1$ is imposed in the
one-dimensional FNLSE with the self-focusing sign of the cubic nonlinearity,
as in that case the model demonstrates the critical and supercritical
collapse at $\alpha =1$ and $\alpha <1$, respectively. However, the collapse
does not take place if the nonlinearity is self-defocusing, which is the
case in Eq. (\ref{FNLSE1}), hence values of $\alpha <1$ may be possible too.

Even and odd functions $-V\left( x\right) $\ and $-W\left( x\right) $ in Eq.
(\ref{FNLSE1}) are the real and imaginary parts of the $\mathcal{PT}$%
-symmetric potential, which represent, respectively, the graded-index
structure and the balanced distribution of the gain and loss \cite{30,31,32}%
. In this work, we adopt them in the natural form of

\begin{equation}
V\left( x\right) =-\frac{1}{2}\left( \frac{x}{w_{0}}\right) ^{2},W\left(
x\right) =-W_{0}x\exp \left( -\frac{x^{2}}{2}\right) ,  \label{PT-Potential}
\end{equation}%
where $w_{0}$ is the width of the corresponding harmonic-oscillator (HO)
potential, and $W_{0}$ is the modulation strength of the gain and loss.
Note that, in the conservative fractional system, with $W(x)=0$, the 
conservation of the integral power, $(d/dz)\int_{-\infty}^{+\infty}\left|\Psi(x,z)\right|^2dx=0$,
is a result of the direct substitution of the expression for $\partial \Psi/\partial z$, taken as
per Eq. (\ref{FNLSE1}).

Setups that realize the nonlinear $\mathcal{PT}$ phase transitions in
optical systems with fractional diffraction were designed on the basis of the Fabry-Perot resonator,
with two convex lenses and two phase masks inserted into it, the fractional diffraction proper being implemented
in the Fourier space \cite{15}. The cubic nonlinearity is incorporated by placing a nonlinear optical material
between the edge mirror and the closest lens. As for the $\mathcal{PT}$ symmetry, its straightforward realization
is provided by adding a balanced pair of gain and loss elements to the resonator \cite{Segev}.

We look for fractional DSs with a real propagation constant $\beta $ as
stationary solutions to Eq. (\ref{FNLSE1}),
\begin{equation}
\Psi \left( z,x\right) =\psi \left( x\right) \exp \left( -i\beta z\right) ,
\end{equation}%
where $\psi \equiv \psi _{R}+i\psi _{I}$ may be split in the real and
imaginary parts. The substitution of this expression into Eq. (\ref{FNLSE1})
yields the stationary equation
\begin{equation}
-\frac{1}{2}\left( -\frac{d^{2}}{dx^{2}}\right) ^{\alpha /2}\psi +\left[
V(x)+iW(x)\right] \psi -|\psi |^{2}\psi +\beta \psi =0.  \label{FNLSE2}
\end{equation}%
Families of DSs are characterized by the integral power (norm), which is a
dynamical invariant of Eq. (\ref{FNLSE1}), defined as
\begin{equation}
P(\beta )=\int_{-\infty }^{+\infty }|\psi |^{2}dx  \label{Power}
\end{equation}%
(formally, it diverges, but in reality it is convergent for a finite-length
spatial domain).

\subsection{The solution method for DSs (dark solitons)}

We seek for fractional DS solutions to Eq. (\ref{FNLSE2}) employing the
Newton conjugate-gradient (NCG) method \cite{33}, with a fixed propagation
constant and soliton power. Accordingly, Eq. (\ref{FNLSE2}) is rewritten as
\begin{equation}
L_{0}\psi =0,  \label{Stationary Equations}
\end{equation}%
where $L_{0}=-\frac{1}{2}\left( -\frac{d^{2}}{dx^{2}}\right) ^{\alpha /2}+%
\left[ V+iW\right] -|\psi |^{2}+\beta $. Here, the boundary condition is $%
\psi =0$ at edges of the spatial domain, which is taken as $-64\leq x\leq 64$%
. As said above, the propagation constant $\beta $ is considered as a
parameter with a fixed value, whereas the solution for $\psi $ is generated
by means of the Newton's iterations. To solve Eq. (\ref{FNLSE2}), we
iteratively update the approximation for $\psi $ as
\begin{equation}
\psi _{n+1}=\psi _{n}+\triangle \psi ,
\end{equation}%
where $\triangle \psi _{n}$ is a small error term. Then we substitute this
expression in Eq. (\ref{FNLSE2}) and expand it around $\psi _{n}$, which
yields
\begin{equation}
L_{0}\psi _{n}+L_{1}\triangle \psi _{n}=O\left( \triangle \psi
_{n}^{2}\right) ,  \label{L0L1}
\end{equation}%
where $L_{1}$ is the linearization operator of Eq. (\ref{Stationary
Equations}) evaluated with the approximate solution $\psi _{n}$. The updated
term $\psi _{n}$ is obtained from the linear Newton-correction equation,
\begin{equation}
L_{1}\triangle \psi _{n}=-L_{0}\psi _{n},
\label{Linear Newton-correction Eq}
\end{equation}%
where
\begin{equation}
L_{1}\triangle \psi _{n}=\left[ -\frac{1}{2}\left( -\frac{d^{2}}{dx^{2}}%
\right) ^{\alpha /2}+V+iW-|\psi |^{2}+\beta \right] \triangle \psi
_{n}-2\psi \text{\textrm{Re}}\left( \psi ^{\ast }\triangle \psi _{n}\right) ,
\label{L1}
\end{equation}%
and $\ast $ stands for the complex conjugate. As $L_{1}$ is not
self-adjoint, we apply to it the adjoint operator of $L_{1}$ and thus turn
it into the following equation:
\begin{equation}
L_{1}^{\dagger }L_{1}\triangle \psi _{n}=-L_{1}^{\dagger }L_{0}\psi _{n}.
\label{Linear Newton-correction Eq-Conj}
\end{equation}%
In the framework of this scheme, Eq. (\ref{Linear Newton-correction Eq-Conj}%
) can be solved directly by dint of conjugate-gradient iterations, which
yields families of DS.

\subsection{The linear stability analysis}

To investigate the stability of the DSs, we apply the standard linearization
procedure, adding a small perturbation to the unperturbed solution:
\begin{equation}
\Psi \left( x,z\right) =e^{-i\beta z}\left[ \psi \left( x\right) +u\left(
x\right) e^{\delta z}+v^{\ast }\left( x\right) e^{\delta ^{\ast }z}\right] ,
\label{Perturbation_psi}
\end{equation}%
where $u(x)$ and $v(x)$ are components of the perturbation eigenmode, with $%
\left\vert u(x)\right\vert ,\left\vert v(x)\right\vert \ll \left\vert \psi
(x)\right\vert $, and $\delta \equiv \delta _{R}+i\delta _{I}$ is the
respective eigenvalue. The stationary solution is stable if all eigenvalues
have $\delta _{R}=0$.

Substituting ansatz (\ref{Perturbation_psi}) in Eq. (\ref{FNLSE1}) and
linearizing it with respect to the perturbation, we arrive at the linear
eigenvalue problem:

\begin{equation}
\delta \cdot u=i\left\{ \left[ -\frac{1}{2}\left( -\frac{d^{2}}{dx^{2}}%
\right) ^{\alpha /2}+V+iW-2|\psi |^{2}+\beta \right] u-\psi ^{2}v\right\} ,
\label{Lineaized-Eqs1}
\end{equation}

\begin{equation}
\delta \cdot v=i\left\{ \psi ^{\ast 2}u+\left[ \frac{1}{2}\left( -\frac{d^{2}%
}{dx^{2}}\right) ^{\alpha /2}-V+iW+2|\psi |^{2}-\beta \right] v\right\} ,
\label{Lineaized-Eqs2}
\end{equation}%
Equations (\ref{Lineaized-Eqs1}) and (\ref{Lineaized-Eqs2}) with the
fractional derivative can be solved by dint of the Fourier collocation
method \cite{33}.

\section{Results and discussion}

\label{Sec III}

\subsection{Families of the fractional DSs and the nonlinear $\mathcal{PT}$
transition}

\begin{figure}[htbp]
\centering\includegraphics[width=0.95\columnwidth]{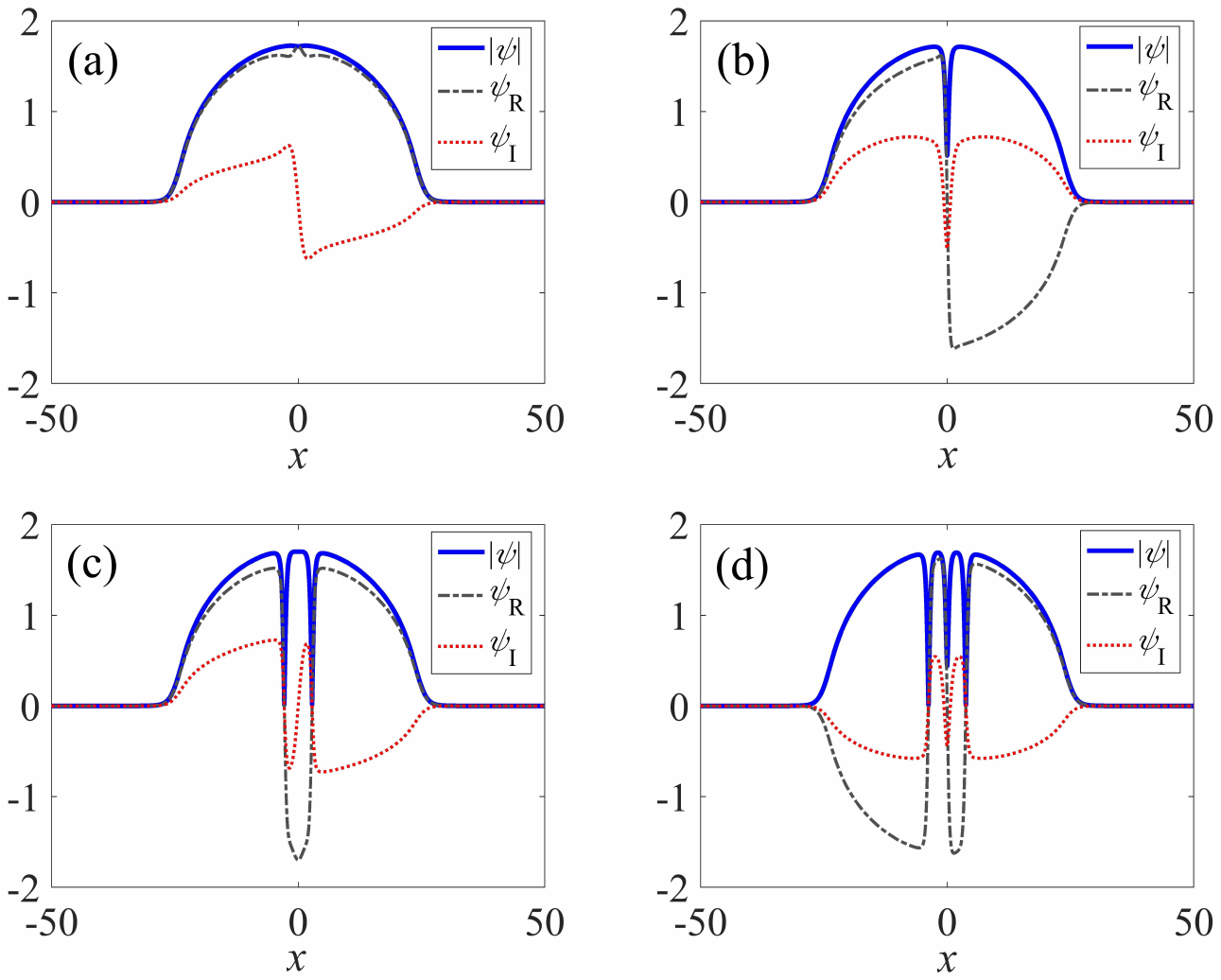}
\caption{Fractional DSs marked by blue-star marks in Fig. \protect\ref%
{figure2}. (a) The fundamental DS (FDS), (b) the 1DS, (c) the 2DS and (d)
the 3DS produced by the numerical solution of Eq. (\protect\ref{FNLSE2}).
Parameters are $\protect\beta =3$, $\protect\alpha =1.7$, $W_{0}=0.2$ and $%
w_{0}=0.1$. All depicted quantities are dimensionless.}
\label{figure1}
\end{figure}

As generic numerically obtained examples, a fundamental DS (FDS) and
high-order (multiple) ones are represented in Fig. \ref{figure1} for a fixed
propagation constant $\beta =3$, LI $\alpha =1.7$, the modulation strength
of the gain-loss distribution $W_{0}=0.2$ and the width of the HO potential $%
w_{0}=0.1$. Figure \ref{figure1}(a) shows the real and imaginary parts of
the fractional FDS, which are even and odd functions of $x$ due to the $%
\mathcal{PT}$ symmetry. The parities of the real and imaginary parts
alternate in the first-, second-, and third-order higher-order fractional
DSs (designated, severally, as 1DS, 2DS, and 3DS modes), as shown in Figs. %
\ref{figure1}(b), (c) and (d), respectively.

\begin{figure}[htbp]
\centering\includegraphics[width=0.95\columnwidth]{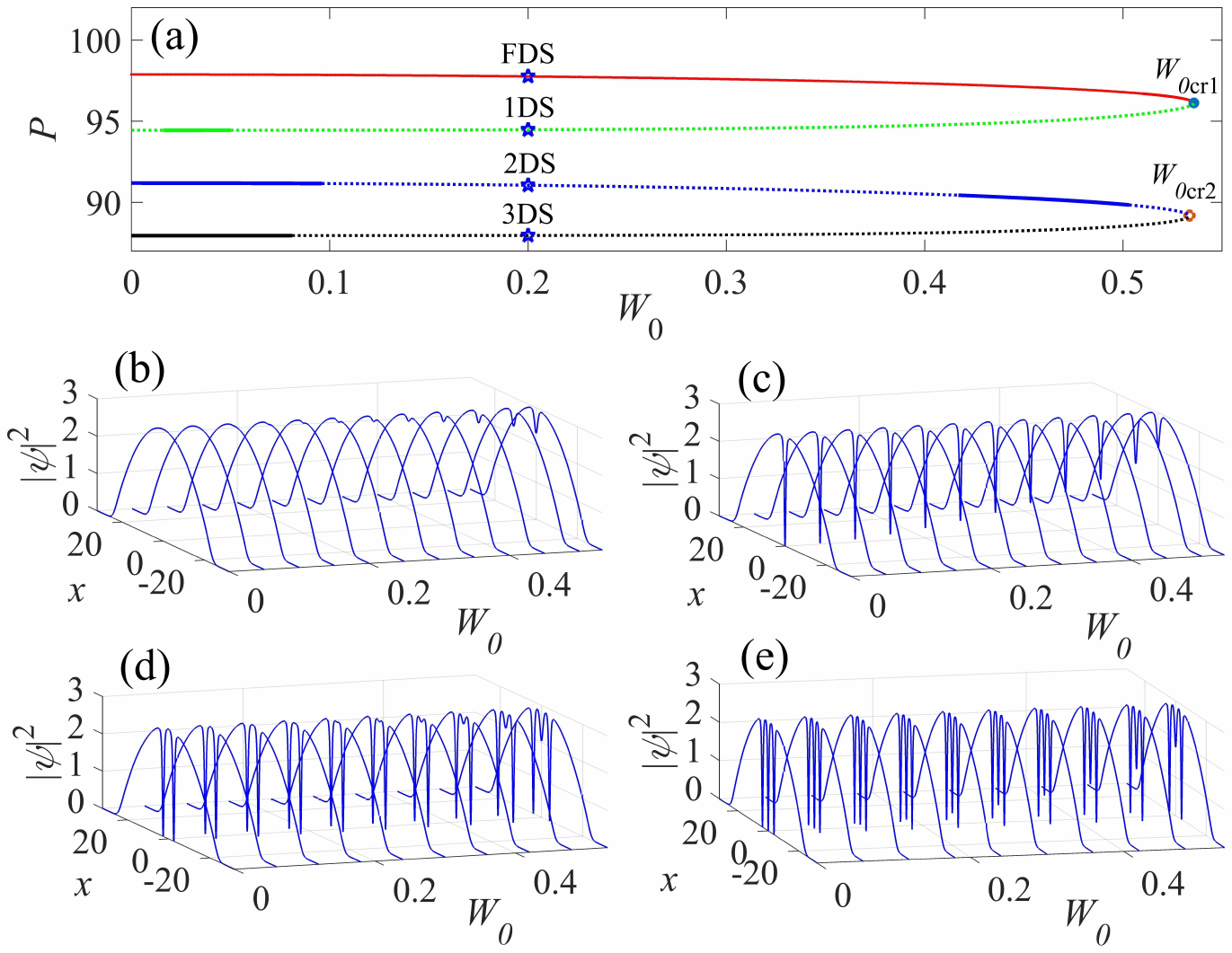}
\caption{The curves of the soliton power $P$\ as the function of the
gain-loss modulation strength $W_{0}$. Solid (dashed) lines indicate stable
(unstable) DS families. The blue and orange dots ($W_{0cr1}$ and $W_{0cr2}$)
indicate the merger points of the FDS-1DS and 2DS-3DS branch pairs. Panels
(b), (c), (d) and (e) display shapes of the FDSs,1DSs, (2DSs and 3DSs,
respectivelyThe parameters are the same as in Fig. \protect\ref{figure1}. }
\label{figure2}
\end{figure}

Next, we consider the effect of the gain-loss modulation strength on the
stability of the fractional DSs. To this end, the numerically found
dependence of the DS's power on the gain-loss modulation strength $W_{0}$ is
shown in Fig. \ref{figure2}(a) for the fixed LI value $\alpha =1.7$. The
diagram encompasses pairwise branches of the DS families, with the pair of
the FDS and 1DS branches merging at the first critical point, $%
W_{0cr1}\approx 0.5357$, and the 2DS -- 3DS pair merging at the second
critical point, $W_{0cr2}\approx 0.5334$. These pairwise mergers represent
nonlinear $\mathcal{PT}$ phase transition in the fractional-diffraction
medium. Further, it is seen that the fractional FDSs are stable in their
entire existence domain, while the stability regions of the higher-order
fractional DSs are piecewise. The dependence of the DS shapes on the
gain-loss modulation strength $W_{0}$ are presented in Figs. \ref{figure2}%
(b)-(e). The shapes of fractional FDSs and 1DSs converge to a similar
profile as $W_{0}$ increases, as shown in Figs. \ref{figure2}(b) and (c),
and eventually become identical at the phase transition points. Actually,
the fractional FDS and 1DS possess the same envelope shapes, but opposite
real and imaginary parts at the phase transition points. A similar scenario
takes place for the 2DS and 3DS branches, as shown in Figs. \ref{figure2}(d)
and (e).

\subsection{The nonlinear $\mathcal{PT}$ phase transition and stability of
the fractional DSs (dark solitons)}

\begin{figure}[htbp]
\centering\includegraphics[width=0.85\columnwidth]{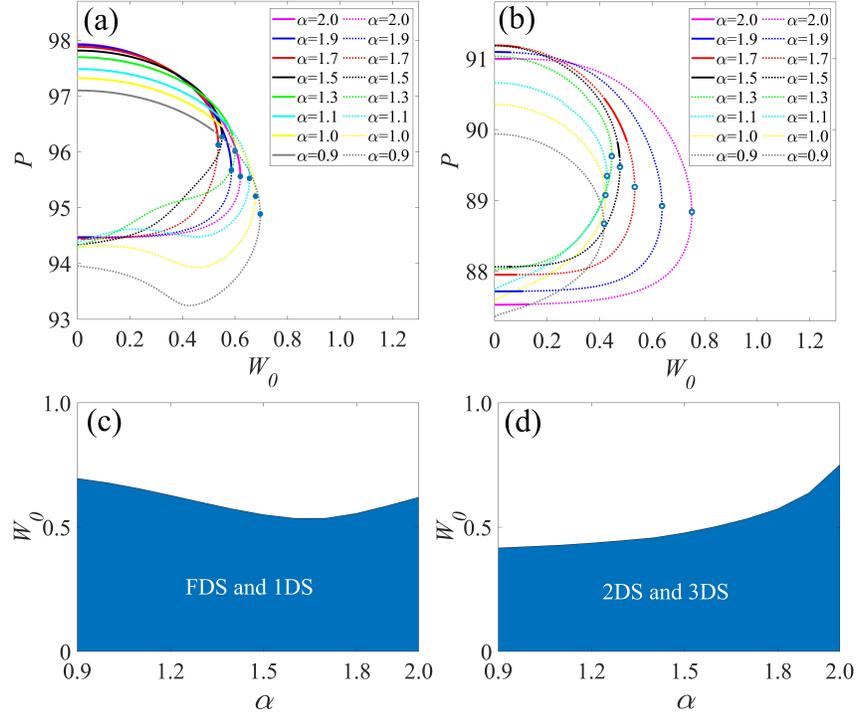}
\caption{The power curves for the FDSs and 1DSs branches (a), and 2DSs and
3DSs ones (b), for different values of LI. Solid (dashed) lines indicate
stable (unstable) families. Blue dots and blue circles indicate the phase
transition points of the FDSs and 1DSs branches, and 2DSs and 3DSs ones,
respectively. Existence regions of the DSs (blue ones) are depicted in panel
(c) for FDSs and 1DSs, and in panel (d) for 2DSs and 3DSs, in the ($W_{0}$,$%
\protect\alpha $) plane. Other parameter are $\protect\beta =3$ and $%
w_{0}=0.1$.}
\label{figure3}
\end{figure}

To investigate the effect of the fractional diffraction on the existence and
stability of DSs, the dependence of the branches of the DS families and
phase transition points on the fractional diffraction is thoroughly analyzed
by means of computing the fractional DS solutions and their stability
spectra on LI in the interval $0.9\leq\alpha \leq 2.0$.

\begin{table}[htbp]
\caption{The critical values of $W_{0}$ at the points of the nonlinear $%
\mathcal{PT}$ phase transitions for one-dimensional DSs.}
\label{table1}\centering
\begin{tabular}{cccccccc}
\hline
{L\'{e}vy index} & $W_{0cr1}$ & $W_{0cr2}$ &  &  & {L\'{e}vy index} & $%
W_{0cr1}$ & $W_{0cr2}$ \\ \hline
$\alpha =2.0$ & $0.6201$ & $0.7509$ &  &  & $\alpha =1.4$ & $0.5730$ & $%
0.4572$ \\
$\alpha =1.9$ & $0.5861$ & $0.6373$ &  &  & $\alpha =1.3$ & $0.6000$ & $%
0.4460$ \\
$\alpha =1.8$ & $0.5554$ & $0.5735$ &  &  & $\alpha =1.2$ & $0.6280$ & $%
0.4360$ \\
$\alpha =1.7$ & $0.5357$ & $0.5334$ &  &  & $\alpha =1.1$ & $0.6548$ & $%
0.4276$ \\
$\alpha =1.6$ & $0.5355$ & $0.5024$ &  &  & $\alpha =1.0$ & $0.6781$ & $%
0.4219$ \\
$\alpha =1.5$ & $0.5502$ & $0.4771$ &  &  & $\alpha =0.9$ & $0.6959$ & $%
0.4164$ \\ \hline
\end{tabular}%
\end{table}

The systematic results for the different values of LI are summarized in Fig. %
\ref{figure3}, in which stable and unstable families are represented by the
solid and dashed segments, respectively, in Figs. \ref{figure3}(a) and (b),
respectively. Figure \ref{figure3}(a) shows that, as above, the fractional
FDSs are stable in their entire existence domain for a relatively weak
fractionality ($\alpha \geq 1.5$), while unstable fractional FDSs occur in a
narrow region near the phase transition points for stronger fractionality ($%
\alpha <1.5$). With the increase of the fractionality, stable regions of the
higher-order DSs shrink and eventually disappear. The most pronounced
difference is observed for 2DSs, which are stable in a narrow region near
the phase transition points for $\alpha =1.5$\ and $\alpha =1.7$ in Fig. \ref%
{figure3}(b).

To quantify the effect of the fractional diffraction on the nonlinear $\mathcal{PT}$ phase transitions, we address the dependence of the 
corresponding critical values of $W_{0}$ on LI. Figures \ref{figure3}(c) and 
\ref{figure3}(d) show $W_{0}$ as a function of LI. Note that the phase-transition points constitute existence boundaries for the fractional DSs. It is noteworthy that, different from the fixed positions of nonlinear $\mathcal{PT}$ phase transitions in non-fractional models \cite{7}, the existence domains of the FDSs and 1DSs first shrink and then gradually expand with the increase of LI, as shown in Fig. \ref{figure3}(c). On the other hand, the existence domain for the 2DSs and 3DSs is monotonically expanding in Fig. \ref{figure3}(d). We identified the boundaries of the existence domains for the fractional DSs by calculating the critical values of $W_{0cr1}$ and $W_{0cr2}$ for the nonlinear $\mathcal{PT}$ phase transitions,  which are summarized in Table \ref{table1} for different LI values, ranging from $0.9$ to $2.0$ with a step of $0.1$.

\subsection{The evolution of the fractional DSs (dark solitons)}

\begin{figure}[htbp]
\centering\includegraphics[width=0.95\columnwidth]{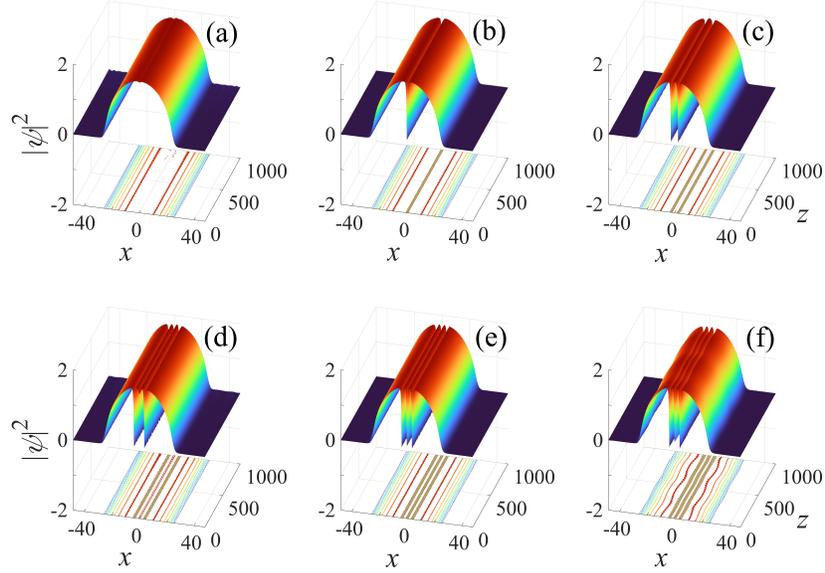}
\caption{The stable evolution of field $\left\vert \protect\psi \right\vert
^{2}$\ of the fractional DSs perturbed by $5\%$ random noise, for LI $%
\protect\alpha =1.7$. (a) The FDS at $W_{0}=0.03$, (b) the 1DS at $%
W_{0}=0.03 $, (c) the 2DS at $W_{0}=0.03$, (d) the 2DS at $W_{0}=0.5$, (e)
the 3DS at $W_{0}=0.03$, (f) the 3DS at $W_{0}=0.08$. Other parameters are $%
\protect\beta =3$ and $w_{0}=0.1$.}
\label{figure4}
\end{figure}

\begin{figure}[htbp]
\centering\includegraphics[width=0.95\columnwidth]{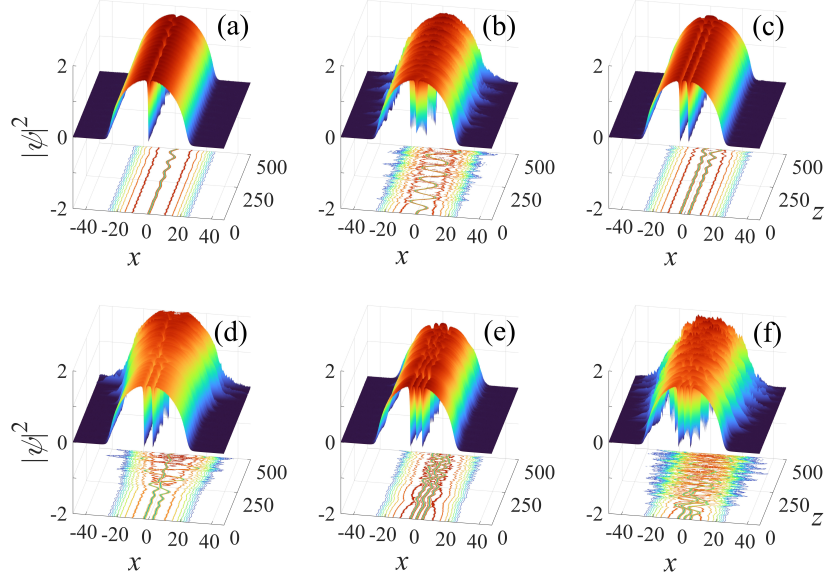}
\caption{The same as in Fig. \protect\ref{figure4}, but for the unstable
evolution with LI $\protect\alpha =1.7$. (a) The 1DS at $W_{0}=0.01$, (b)
the 1DS at $W_{0}=0.1$, (c) the 2DS at $W_{0}=0.12$, (d) the 2DS at $%
W_{0}=0.15$, (e)\ the 3DS at $W_{0}=0.09$, (f) the 3DS at $W_{0}=0.15$.
Other parameters are $\protect\beta =3$ and $w_{0}=0.1$.}
\label{figure5}
\end{figure}

To corroborate the predictions of the linear-stability analysis, we have
performed numerical simulations of Eq. of the perturbed evolution of the
fractional DSs of Eq. (\ref{FNLSE1}). Typical examples are presented in
Figs. \ref{figure4} and \ref{figure5} for fixed values of $\alpha =1.7$, $%
\beta =3$, and $w_{0}=0.1$.

Representative examples of possible stable dynamics of the fractional FDSs
and high-order dark solitons are shown in Fig. \ref{figure4}. Linearly
stable FDSs and 1DSs are perturbed by $5\%$ random noise for $\alpha =1.7$
at $W_{0}=0.03$, as shown in Figs. \ref{figure4}(a) and (b). The results
confirm that the linearly stable FDS and 1DS, as predicted by the
linear-stability analysis, remain robust, at least, up to $z=1000$. The
linearly stable 2DSs exist in two different regions, as shown in Fig. \ref%
{figure2}(a), therefore we separately examine the stability and evolution of
the 2DSs at $W_{0}=0.03$ and $W_{0}=0.5$. It is seen in Figs. \ref{figure4}%
(c) and (d) that, at least up to $z=1000$, the linearly stable 2DSs indeed
remain stable in the course of the perturbed propagation. The evolution
results displayed in Figs. \ref{figure4}(e) and \ref{figure4}(f) confirm
that, under the $5\%$ noise perturbation, the stable fractional 3DSs also
survive as stable states.

To demonstrate the unstable evolution, fractional DSs, which are predicted
to be unstable by the linear-stability analysis, are also disturbed by $5\%$
random noise. Figure \ref{figure5} displays typical examples of the
evolution of unstable 1DSs [panels Figs. \ref{figure5}(a) and (b)], 2DSs
[panels Figs. \ref{figure5}(c) and (d)], and 3DSs [panels Figs. \ref{figure5}%
(e) and (f)]. The evolution of the unstable fractional DSs demonstrates
spontaneous onset of oscillations and displacement of the center.

\subsection{The two-dimensional system}

\begin{figure}[htbp]
\centering\includegraphics[width=0.9\columnwidth]{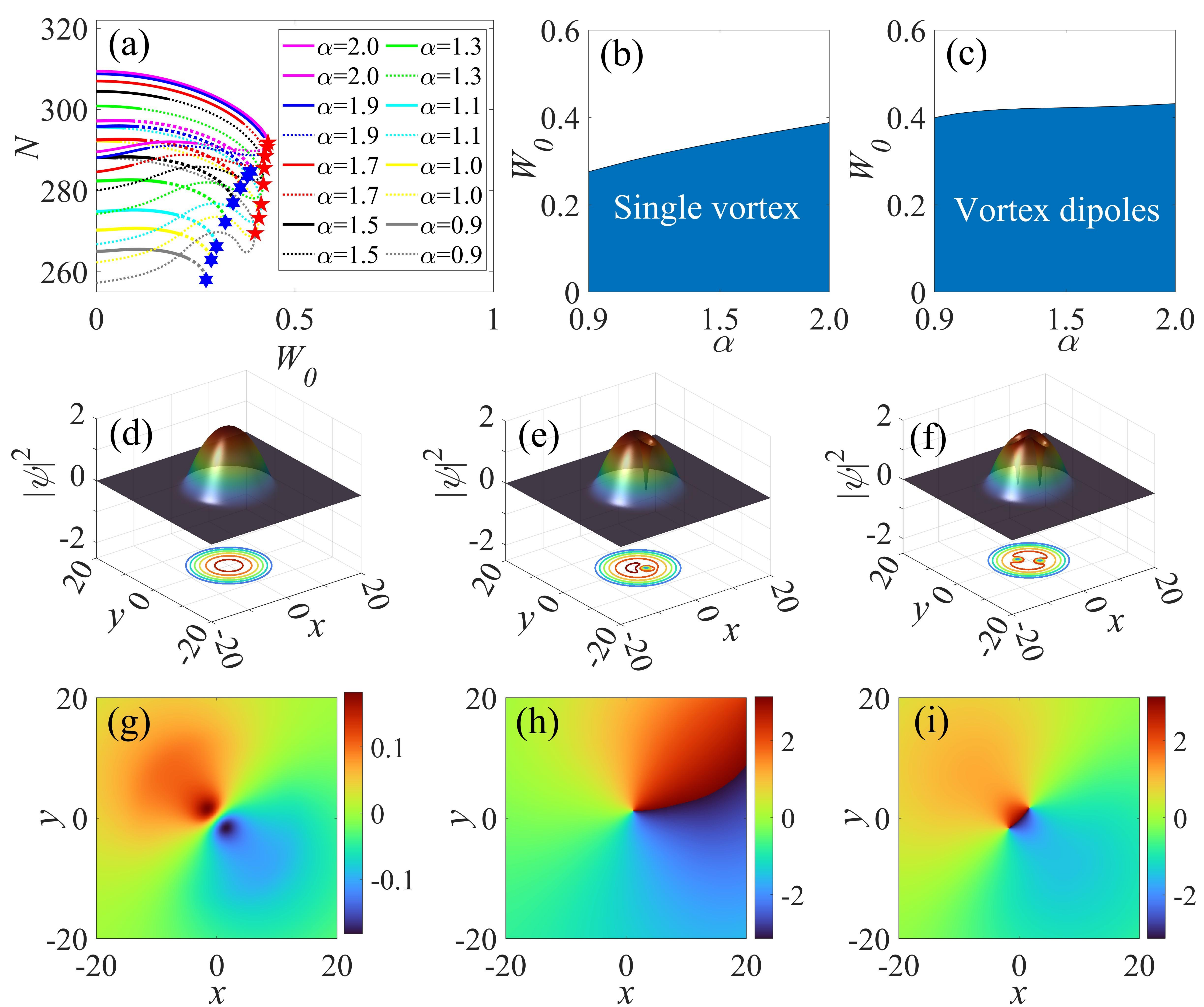}
\caption{The power curves $P\left( W_{0}\right) $ for the branches of the
two-dimensional FDSs, single vortex and vortex-dipole families (a). Solid
(dashed) lines indicate stable (unstable) families, with red pentagram and
blue hexagon indicating the respective phase transition points. The blue
domain in panels (b) and (c) represent the existence region of the
two-dimensional FDSs, single vortices, and the vortex dipoles in the ($W_{0}$,$%
\protect\alpha $) plane. Examples of the stable fractional two-dimensional
FDS, single vortex and vortex dipole are displayed in panels (d), (e) and
(f), respectively, for $\protect\beta =2$, $W_{0}=0.05$ and $\protect\alpha =1.7$. Panels
(g), (h) and (i) show the phase patterns of the two-dimensional FDS, single
vortex and the vortex dipole, respectively.}
\label{figure6}
\end{figure}

Finally, we consider the two-dimensional $\mathcal{PT}$-symmetric potential
with a real part
\begin{equation}
V\left( x,y\right) =-\frac{1}{2}\left( \frac{x^{2}+y^{2}}{w_{1}^{2}}\right) ,
\label{PT-2D-Re}
\end{equation}%
which represents the two-dimensional HO trap, and the odd imaginary part

\begin{equation}
W\left( x,y\right) =-W_{0}\left( x+y\right) \exp \left( -\frac{x^{2}+y^{2}}{4%
}\right) .  \label{PT-2D-Im}
\end{equation}

To investigate effects of the fractional diffraction on the nonlinear $%
\mathcal{PT}$ phase transitions and stability of two-dimensional DSs, the
dependence of branches of two-dimensional FDSs, single-vortex, and
vortex-dipole families on the diffraction fractionality are analyzed by
computing the solutions and respective stability spectra for a fixed HO
width, $w_{1}=5$, and LI ranging from $0.9$ to $2.0$. Here, the boundary
condition is $\psi =0$ at edges of the spatial domain, which is taken as $%
-32\leq x,y\leq 32$.

The power curves for the families of two-dimensional FDS, single vortex and
vortex dipole (i.e., vortex-antivortex bound states) are plotted in Fig. \ref%
{figure6}(a), where the branches disappear at the respective critical values
of $W_{0}$. Note that the two-dimensional FDSs are stable in their entire
existence domains of existence solely for $\alpha =2.0$ (the usual NLSE with
the non-fractional diffraction). In the case of the fractional diffraction ($%
\alpha <2$) instability of two-dimensional FDSs occurs. Thus, the stability
range for the two-dimensional FDSs gradually shrinks with the increase of
the fractionality (decrease of LI). In contrast, the stability range for the
single vortex gradually expands with the decrease of LI. Vortex dipoles are
stable only in narrow subregions of the their existence domain and become
completely unstable as $\alpha \leq 1.5$.

\begin{table}[htbp]
\caption{The critical values of the gain-loss modulation strength, $W_{0}$,
at the nonlinear $\mathcal{PT}$ phase transition for single vortex and
vortex dipoles.}
\label{table2}\centering
\begin{tabular}{cccccccc}
\hline
{L\'{e}vy index} & $W_{0cr3}$ & $W_{0cr4}$ &  &  & {L\'{e}vy index} & $%
W_{0cr3}$ & $W_{0cr4}$ \\ \hline
$\alpha =2.0$ & $0.3885$ & $0.4319$ &  &  & $\alpha =1.4$ & $0.3344$ & $%
0.4216$ \\
$\alpha =1.9$ & $0.3800$ & $0.4293$ &  &  & $\alpha =1.3$ & $0.3243$ & $%
0.4203$ \\
$\alpha =1.8$ & $0.3714$ & $0.4271$ &  &  & $\alpha =1.2$ & $0.3136$ & $%
0.4182$ \\
$\alpha =1.7$ & $0.3625$ & $0.4252$ &  &  & $\alpha =1.1$ & $0.3021$ & $%
0.4148$ \\
$\alpha =1.6$ & $0.3535$ & $0.4238$ &  &  & $\alpha =1.0$ & $0.2890$ & $%
0.4090$ \\
$\alpha =1.5$ & $0.3441$ & $0.4226$ &  &  & $\alpha =0.9$ & $0.2759$ & $%
0.4000$ \\ \hline
\end{tabular}%
\end{table}

The boundaries of the existence domain of the two-dimensional FDSs and
vortex dipoles, as well as the the single vortex are obtained by calculating
the respective critical values $W_{0cr3}$ and $W_{0cr4}$ at the nonlinear $%
\mathcal{PT}$ phase transition points. These critical values are summarized
in Table \ref{table2}, which demonstrates that the critical values of $W_{0}$
decrease as the fractionality increases. This means that the existence
domains of the two-dimensional FDSs, single vortices, and vortex dipoles shrink
with the decrease of LI, as shown in Fig. \ref{figure6}(b) and (c).
Representative examples of stable two-dimensional FDS, single vortex and
vortex dipole are displayed, respectively, in Figs. \ref{figure6}(d), (e)
and (f), along with their phase profiles in Figs. \ref{figure6}(g), (h) and
(i).

\begin{figure}[htbp]
\centering\includegraphics[width=0.75\columnwidth]{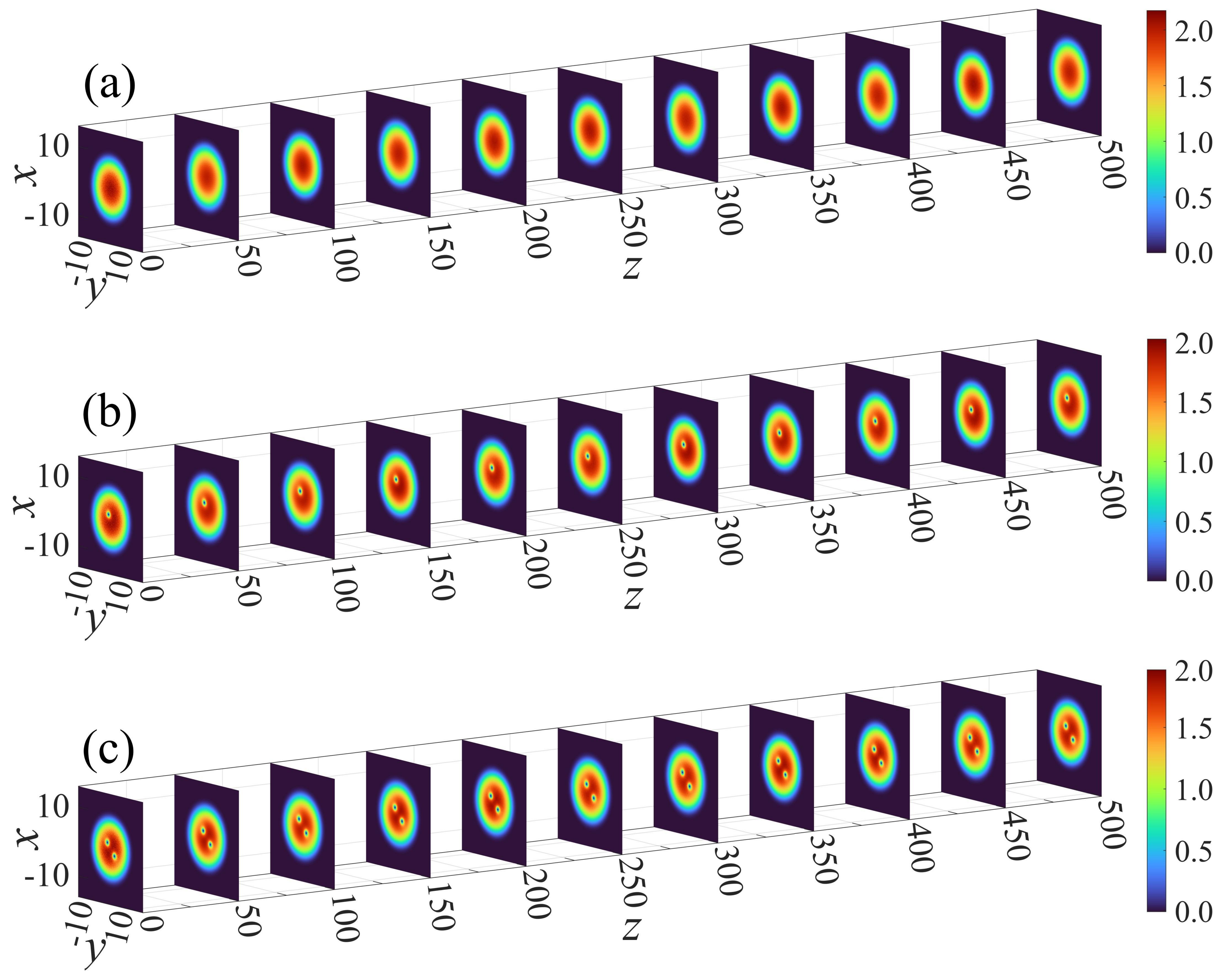}
\caption{Stable evolutions of the two-dimensional FDS (a), single vortex (b),
and vortex dipole (c) for $\protect\beta =2$, $W_{0}=0.05$ for $\protect%
\alpha =1.7$ and random perturbations at a 5$\%$ amplitude level.}
\label{figure7}
\end{figure}

\begin{figure}[htbp]
\centering\includegraphics[width=0.75\columnwidth]{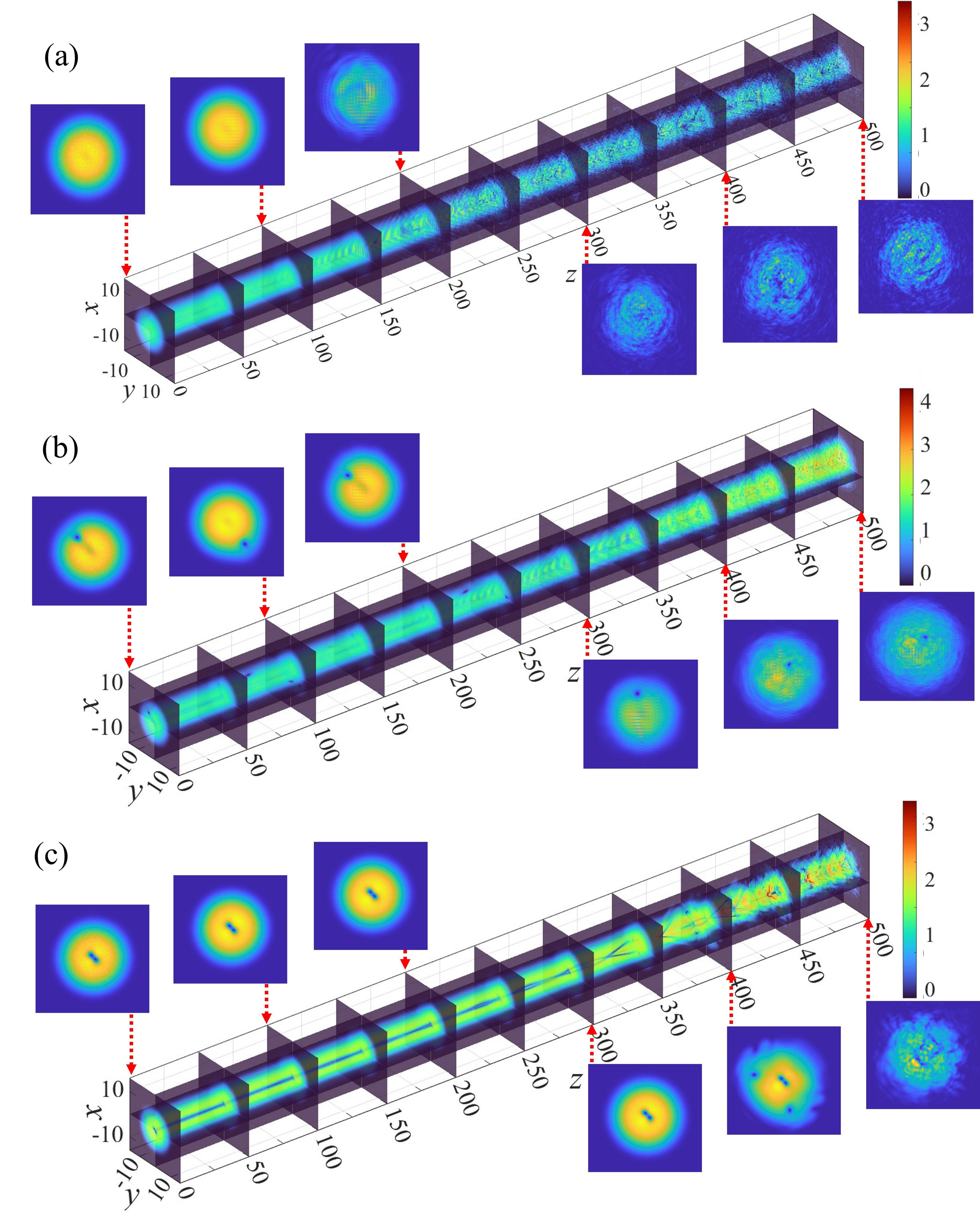}
\caption{The unstable evolution of the two-dimensional FDS (a), single vortex
(b) and vortex dipole (c) with $\protect\beta =2$, $W_{0}=0.3$ for $\protect%
\alpha =1.5$.}
\label{figure8}
\end{figure}

Next, we examine the stability and evolution of the two-dimensional FDSs,
single vortices, and vortex dipoles by means of direct simulations of their
perturbed evolutions. The robust propagation of the two-dimensional FDS at
parameters corresponding to the stable region identified in Fig. \ref%
{figure6}(a) is indeed demonstrated by the simulations performed in the
presence of perturbations. As shown in Fig. \ref{figure7}(a), the perturbed
propagation remains stable at least up to $z=500$ under the action of random
initial perturbations with a relative amplitude of $5\%$. Figures \ref%
{figure7}(b) and (c) corroborate the robust propagation of the stable single
vortex and vortex dipole.

On the other hand, examples of the unstable evolution of the two-dimensional
FDS, single vortex, and vortex dipole are presented in Fig. \ref{figure8}.
The simulations, their instability, as predicted by the linear-stability
analysis (the onset of the instability, leading to emergence of
\textquotedblleft turbulence", is readily observed even in the absence of
random perturbations). Figure \ref{figure8} (a) shows that the unstable
two-dimensional FDS survives as a quasi-stable mode at the early stage of
the evolution ($z<100$), which is followed by the onset of conspicuous
instability. Figs. \ref{figure8}(b) and (c) show that the unstable vortex
dipole survives as a quasi-stable mode up to $z=200$; then, it starts to
rotate and eventually collapses.

\section{Conclusion}

\label{Sec IV} In the present work, we have produced numerical solutions for
the one- and two-dimensional fractional FDSs (fundamental-dark-soliton), as
well as one-dimensional higher-order DS (dark-soliton), two-dimensional
single-vortex, and vortex-dipole families. The analysis reveals the nonlinear
$\mathcal{PT}$ phase transitions (the merger of pairs of branches
representing different families), which suggests possibility for new
experiments in fractional nonlinear optical systems. The fractional
diffraction produces a significant effect on the nonlinear $\mathcal{PT}$
phase transition, reflected in the dependence of parameters of the
respective merger points on the LI (L\'{e}vy index). The existence and
stability regions for the fractional DSs, single vortices, and vortex dipoles
have been identified. Direct simulations of the
perturbed evolution of these modes fully agree with the predictions of the
linear-stability analysis. The complex stability patterns of the fractional
DSs and vortex dipoles uncover multiple effects of the fractional
diffraction, $\mathcal{PT}$ symmetry, and nonlinearity.

Extension of the present analysis may address vortex solitons with higher
topological charges, vortex rings and other forms of two-dimensional DS
structures.

\begin{backmatter}
\bmsection{Funding}
National Natural Science Foundation of China (11805141); Shanxi Province Basic Research Program (202203021222250, 202303021211185); Israel Science Foundation (grant No. 1695/22).

\bmsection{Disclosures}
The authors declare no conflicts of interest.

\bmsection{Data Availability Statement}
Data underlying the results presented in this paper are not publicly available at this time but may be obtained from the authors upon reasonable request.
\end{backmatter}

\end{document}